\documentclass[manuscript,nofootinbib,noshowkeys,superscriptaddress]{revtex4}
\usepackage[latin1]{inputenc}
\usepackage{pstricks,pst-node,pst-text,pst-3d,pslatex}
\usepackage[T1]{fontenc}

\usepackage{amssymb,amsfonts,amsmath}
\usepackage{graphicx}

\newcommand{\be}{\begin{eqnarray}}
\newcommand{\ee}{\end{eqnarray}}

\begin{document}


\title{Dominant Folding Pathways of a WW Domain}

\author{Silvio a Beccara}
\affiliation{Dipartimento di Fisica  Universit\`a degli Studi di Trento, Via Sommarive 14, Povo (Trento), I-38123 Italy}
\affiliation{INFN, Gruppo Collegato di Trento, Via Sommarive 14, Povo (Trento), I-38123 Italy.},
\author{Tatjana \v{S}krbi\'c}
 \affiliation{European Centre for Theoretical Studies in Nuclear Physics and
  Related Areas, Strada delle Tabarelle 286, Villazzano (Trento), I-38123 Italy}
\author{Roberto Covino}
\affiliation{Dipartimento di Fisica  Universit\`a degli Studi di Trento, Via Sommarive 14, Povo (Trento), I-38123 Italy}
\affiliation{INFN, Gruppo Collegato di Trento, Via Sommarive 14, Povo (Trento), I-38123 Italy.}
\author{Pietro Faccioli~\email{Corresponding author: faccioli@science.unitn.it}}
\affiliation{Dipartimento di Fisica  Universit\`a degli Studi di Trento, Via Sommarive 14, Povo (Trento), I-38123 Italy}
\affiliation{INFN, Gruppo Collegato di Trento, Via Sommarive 14, Povo (Trento), I-38123 Italy.}

\begin{abstract}
We investigate the folding mechanism of the WW domain Fip35 using a realistic atomistic force field
by applying the Dominant Reaction Pathways (DRP) approach.
We find evidence for the existence of two folding pathways, which differ by the order of formation of the two hairpins. 
This result is consistent  with the analysis of the experimental data on the folding kinetics of WW domains and with the results obtained from large-scale molecular dynamics (MD) simulations of this system.  Free-energy calculations performed in two coarse-grained models support the 
robustness of our results and suggest that the qualitative structure of the dominant paths are mostly shaped by the native interactions. 
Computing a folding trajectory in atomistic detail only required about one hour on 48 CPU's.
The gain in computational efficiency opens the door to a systematic investigation of the folding pathways of a large number of globular proteins.
\end{abstract}

\maketitle
Unveiling  the mechanism by which proteins fold into their
native structure remains one of the fundamental open problems at the interface of contemporary molecular biology, biochemistry and biophysics.
A critical point concerns the characterization of the 
ensemble of reactive trajectories connecting the denatured and native states, in configuration space.  

In this context, a fundamental question which has long been debated~\cite{foldingtheory1} is whether the folding of typical globular proteins involves 
a few dominant pathways, i.e.  well defined and conserved sequences
of secondary and tertiary contact formation, or if it can take place through a multitude of qualitatively different 
routes. A related important question  concerns the role of non-native interactions in determining the structure of the folding 
pathways~\cite{foldingtheory2, foldingtheory3}. 

In principle, atomistic MD simulations provide a consistent framework to address these problems from a theoretical perspective. 
However, due to their high computational cost,  MD simulations can presently only be used to investigate the conformational dynamics of relatively small
 polypeptide chains,  and are only able to cover time intervals much smaller than the  folding times of typical globular proteins.

In view of these limitations, a considerable amount of  theoretical and experimental activity has been devoted to investigate the 
folding of protein sub-domains, which consist of only a few tens of aminoacids, and fold on sub-millisecond time-scales~\cite{expfastfolders}. 
 In particular, a number of mutants of the 35 aminoacid WW domain of human protein pin1   
 have been engineered which fold in few tens  of microseconds ~\cite{expWW1}. 
Their small size and their ultra-fast kinetics make them ideal benchmark systems, for which numerical simulations
can be compared with a large body of experimental data~\cite{expWW1, expWW2, expWW3}. 

\begin{figure}
\includegraphics[width=10 cm]{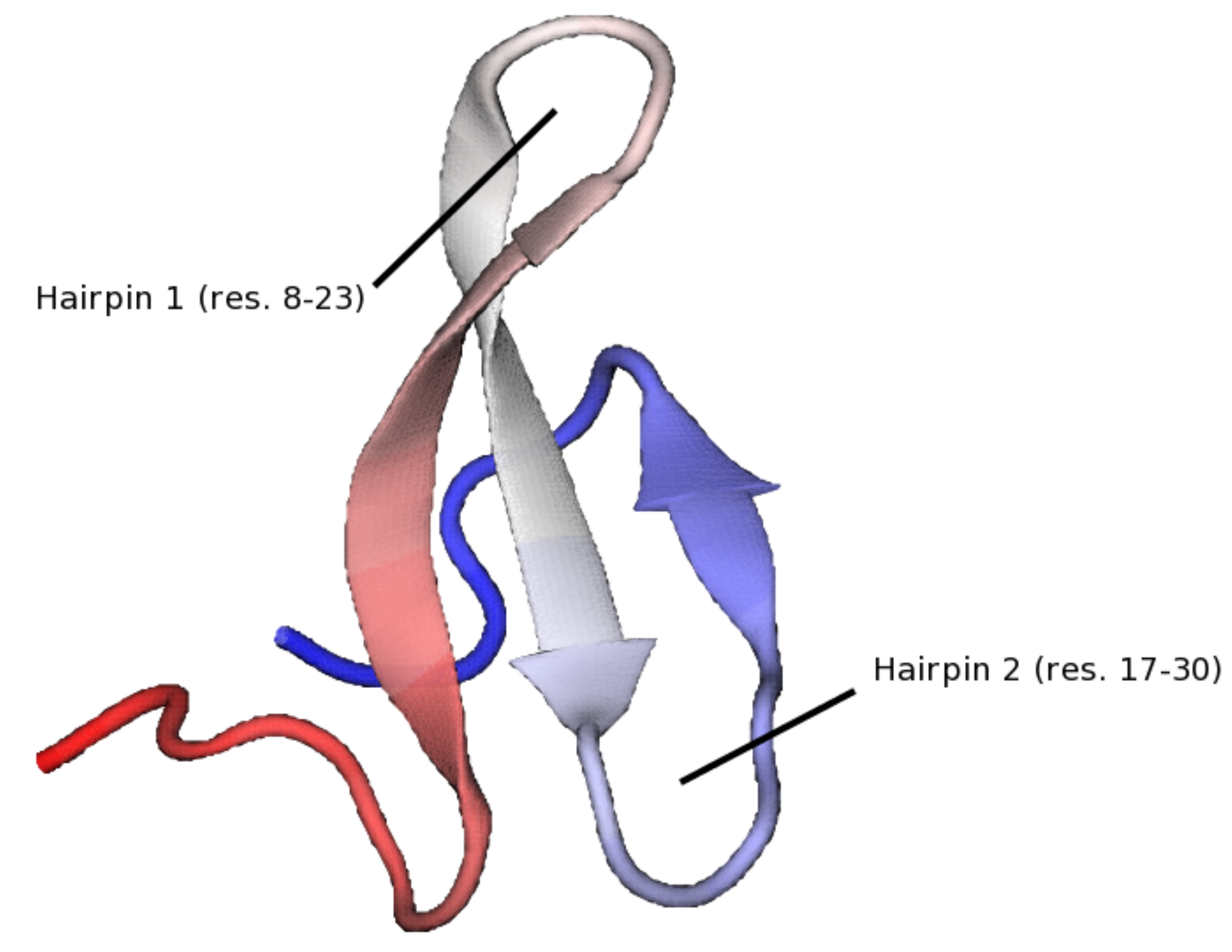}
\caption{
{\bf Native structure of Fip35, a WW domain of the fip mutant of protein human pin1 (pdb code: pin1).  The primary sequence of fip35 is:
EEKLPPGWEKRMSADGRVYYFNHITNASQWERPSG. }}
\label{Fig1}
\end{figure}

In  particular, a MD simulation was performed to investigate the dynamics of a mutant named Fip35 (see Fig. \ref{Fig1}),  
for a time interval longer than $10\;\mu$s. Unfortunately, in that simulation
no folding transition was observed \cite{theWW1, theWW2}.      

The folding of this WW domain was later investigated by Pande and co-workers, using a world-wide distributed computing scheme ~\cite{theWW3}. 
According to this study the transition proceeds in a very heterogeneous way, 
i.e. through a multitude of qualitatively different and nearly equiprobable folding pathways. 

\begin{figure}
\includegraphics[width=14 cm]{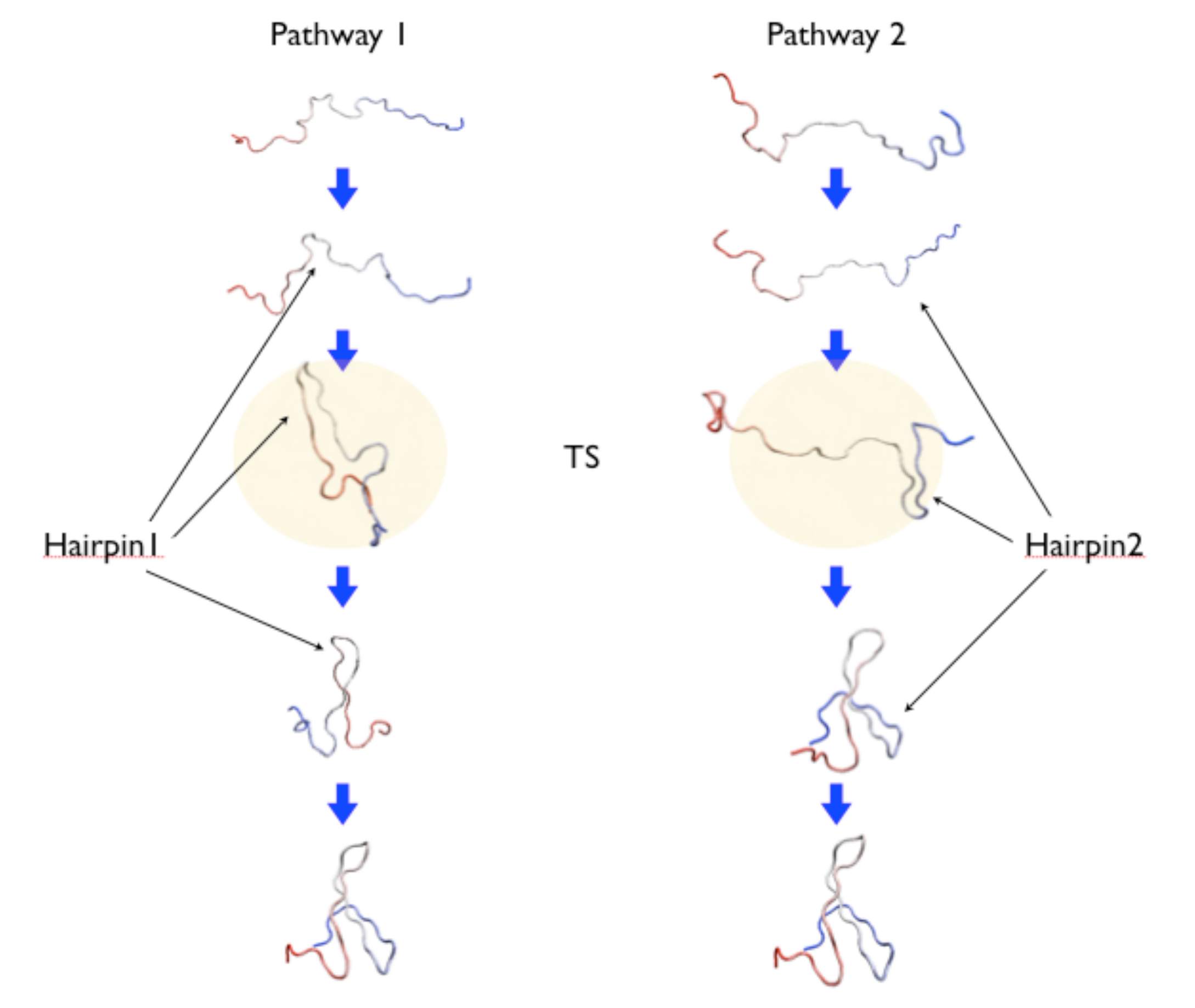}
\caption{\bf 
Schematic representation of the structure of the two folding pathways obtained in our DRP simulations. }
\label{Fig2}
\end{figure}

No\'e and co-workers performed a Markov state model analysis of a large number of short ($\lesssim$~200~ns)  
nonequilibrium MD trajectories~\cite{theWW4} performed on the WW domain of human Pin 1 protein. 
They reported a complex network of transition pathways, which differ by the specific order in which the different local meta-stable states were visited.
On the other hand, in all pathways the formation of hairpins takes place in a definite sequence (see e.g. Fig. \ref{Fig2}). 
In particular, from the statistical model it was inferred that  in about $30\%$ of the folding transitions,  the second hairpin forms first, as in the right panel.    

A different conclusion has been reached by Shaw and co-workers, by analyzing a ms-long MD trajectory with multiple unfolding/refolding events, obtained using a 
special-purpose supercomputer \cite{theWW5}. In that simulation the WW domain of Fip35 was found to fold and unfold predominantly along a pathway in which 
hairpin 1 is fully structured, before hairpin 2 begins to fold, as shown in the left panel of Fig.\ref{Fig2}.  
In a recent paper~\cite{krivov}, Krivov re-analyzed the same ms-long MD trajectory in order to
identify an optimal set of reaction coordinates.
His conclusion was that the folding of this WW domain is thermally activated rather than incipient downhill and that the transition also occurs through a second pathway, in which hairpin 2 forms before hairpin 1. The statistical weights of the two pathways estimated from the number of folding events are  $80\%\pm 20\%$ and $20\%\pm 10\%$.  

While all these theoretical studies yield folding times in rather good agreement with available experimental data on folding kinetics,  
they provide different pictures of the folding mechanism and raise a number of issues.

Firstly, it is important to assess the degree of heterogeneity of the folding mechanism and to clarify whether the most statistically significant folding pathways are those in which the hairpins form in sequence. 
Important related questions are also whether the folding mechanism is correlated with the structure of the initial denatured conditions from which the reaction is initiated and
 with the temperature of the heat bath.
Finally, it is interesting to address the problem of  the relative role played by native and non-native interactions in determining the structure of folding pathways. Indeed, while native interactions are arguably shaping the dynamics
in the vicinity of the native state, non-native interactions may in principle play an important role in the transition region and at the rate limiting stages of the reaction. 

In order to tackle these questions, in this work we use the Dominant Reaction Pathways (DRP) approach \cite{DRP1,DRP2,DRP3, DRP4, QDRP1}, a framework which allows to very efficiently compute the statistically  most  significant pathways connecting given denatured configurations to the native state at an atomistic level of detail, with realistic force fields.  To further support our results and to study the role of native and non-native interactions we map the free energy landscape by performing Monte Carlo simulations in two coarse-grained models.  
 
 Our study confirms that Fip35 folds mostly through the two folding channels discussed above and shows that the relative weight of the two channels changes with
 temperature. In addition, we find that the folding pathways are  correlated with the initial condition from which the transition is initiated.
 The studies based on the coarse-grained model suggest that the folding dynamics in the transition region 
 is not significantly influenced by non-native interactions.
 
\section{Methods}

  \subsection{Atomistic Force Field}

Our atomistic simulations of the dominant folding trajectories of the Fip35 WW domain were performed using the AMBER~ff99SB force field~\cite{AMBER99sb} in implicit solvent with
Generalized Born formalism implemented in GROMACS 4.5.2 \cite{GRO4}. The Born radii were calculated according to the Onufriev-Bashford-Case algorithm \cite{OBC}.

In a recent work based on the DRP method, the dominant pathway in the conformational transition of tetra-alanine obtained using the same version of the AMBER 
force field was found to agree well with the results of an analogous calculation in which the 
molecular potential energy  was determined {\it ab initio}, i.e. directly from quantum  electronic structure calculations~\cite{QDRP2}. 

\subsection{Coarse-grained Model}

To study the equilibrium properties of the folding of the Fip35 WW domain we used the coarse-grained model recently developed in Ref.s~\cite{Kim&Hummer, 
reactioncoord2}. In that model, 
aminoacids are represented by spherical beads centered at the $C_\alpha$ positions. The non-bonded part of the potential energy contains both native and non-native interactions. 
 The former are the same used in the G$\overline{o}$-type model  of Ref.~\cite{Karanikolas&Brooks}, while the latter consist of a quasi-chemical potential,
  which accounts for the  statistical propensity of different aminoacids to be found in contact in native structures, and of a Debye-screened electrostatic term. 
  In this model, the
   average potential energy due to native interactions in the folded phase is typically one order of magnitude larger than that due to non-native interactions. 
    Above the folding temperature,    this ratio drops to about 4. 
 
   This model was shown to provide an accurate description of protein-protein complexes  with low and intermediate binding affinities \cite{Kim&Hummer}. In the
   insert of the first panel of Fig. \ref{Fig3} we plot the specific heat, evaluated from MC simulations at different temperatures, which indicates that this model
yields the correct folding temperature for this WW domain.
  
\subsection{The Dominant Reaction Pathways Method}
 
The high computational cost of MD simulations of macromolecular systems has triggered efforts towards developing alternative
 theoretical frameworks to investigate their long-time dynamics and reaction kinetics (see e.g. \cite{DRP1, method1, method2, method3, method4, method5, method6} and references therein). 

In particular, the DRP approach \cite{DRP1, DRP2, DRP3,DRP4, QDRP1, DRP0} concerns physical systems which can be described by the overdamped Langevin equation. If ${\bf x}_k$ denotes the
coordinate of the $k-$th atom, the Langevin equation in the so-called Ito Calculus reads: 
\begin{equation} \label{langev_eq}
{\bf x}_k(i+1)= {\bf x}_k(i) -\frac{\Delta t D_k}{k_B T} {\bf \nabla}U[{\bf X}(i)] + \sqrt{2 D_k \Delta t}\, {\bf \eta}_k(i).
\end{equation} 
In this equation, ${\bf X}(i)\equiv ({\bf x}_1(i),\ldots, {\bf x}_N(i))$ is the set of atomic coordinates at the $i-$th time step, $\Delta t$ is an elementary time interval, $D_k$ is the diffusion 
coefficient of the $k-$th atom, $k_B$ is the Boltzmann's constant, $T$ is the temperature of the heat-bath,  
$U({\bf X})$ is the  potential energy. ${\bf \eta}^k(i)$ is a white Gaussian noise with unitary variance, acting on the $k-$th atom. 

The probability for a protein to fold in a given time interval $t$  can be written as
\be 
\label{Pf}
 P_f(t) = \int d{\bf X}_f~h_N({\bf X}_f)  \int d {\bf X}_i~h_D({\bf X}_i)~P({\bf X}_f, t| {\bf X}_i)~\rho_0({\bf X}_i), 
 \ee
where $h_{N(D)}({\bf X})$ is the characteristic function of the native (denatured) state (defined in terms of some suitable order parameters), $\rho_0({\bf X}_i)$ is the initial distribution of 
micro-states in the denatured state  and $P({\bf X}_f, t|{\bf X}_i)$ is the conditional probability of reaching the (native)
configuration ${\bf X}_f$ starting from the (denatured) configuration ${\bf X}_i$, in a time $t$.  If the total time interval $t$ is chosen much smaller than the inverse folding rate,
 this probability  is dominated by single nonequilibrium folding events.  

It can be shown that the probability of a given folding trajectory ${\bf X}(t)$ connecting denatured and native configurations  is proportional to the negative exponent of the Onsager-Machlup functional \cite{DRP0, method5}, which in discretized form reads
\begin{eqnarray}
&&\text{Prob}[{\bf X}] \propto \exp\left[-\sum_{i=1}^{N_t} \sum_{k=1}^N
\frac{1}{4 D_k \Delta t} \right. \nonumber\\
&&\cdot \left.  \left({\bf x}^k(i+1)-{\bf x}^k(i) + \frac{\Delta t ~D_k}{k_B T} \nabla U[{\bf X}(i)] \right)^2\right],
\label{prob}
\end{eqnarray}
where $N_t$ is the number of time steps in the trajectory. On the other hand, the paths which do not reach  native state before time $t$ do not contribute to the
 transition probability in Eq.~{\bf \ref{Pf}}. 
The most probable ---or so-called \emph{dominant}--- reaction pathways are those which minimize the exponent in Eq. {\bf \ref{prob}}. In principle, these may be found by numerically
relaxing the effective action functional~\cite{DRP1, Adib}
\be
S_{eff}[{\bf X}] = \Delta t~\sum_{i=1}^{N_t}~\left[\sum_{k=1}^N \frac{({\bf x}_k(i+1)-{\bf x}_k(i))^2}{4 D_k \Delta t^2} + V_{eff}[{\bf X}(i)]\right], 
\label{Seff}
\ee
where $V_{eff} ({\bf X})$ is the so-called effective potential, and reads
\begin{equation} 
 V_{eff} ({\bf X}) = \frac{1}{4 (k_B T)^2} \sum_k  D_k~\left(~ \left| {\bf \nabla}_k~U({\bf X})  \right|^2 - 2~k_B
T\; \nabla_k^2~U({\bf X}) \ \right).
\label{Veff}
\end{equation} 

In practice, for a protein folding transition, directly minimizing the effective action in Eq.~{\bf \ref{Seff}} is unfeasible, since at least $10^4-10^5$ time steps are needed to describe a single 
folding event.   On the other hand, for any fixed pair of native and denatured configurations the dominant paths can be equivalently  found by 
minimizing an effective Hamilton-Jacobi (HJ) action in the form~\cite{DRP1,DRP0}
\begin{equation}
S_{HJ} =\sum_{i=1}~\Delta l_{i,i+1} \sqrt{\frac{1}{D}\left(E_{eff} + V_{eff}[{\bf X}(i)]\right)},
\label{SHJ}
\end{equation}
where and $\Delta l_{i+1, i}=\sqrt{({\bf X}(i+1)-{\bf X}(i))^2}$ represents the elementary 
displacement in configuration space, and for sake of clarity, we have assumed that all atoms have the same diffusion coefficient.  
The parameter $E_{eff}$ determines the  time at which any given frame $l$ of the path is visited:   
 \be
 t(l) = \sum_{i\le l} \Delta l_{i, i+1} ~\hspace{0.1cm}[4 D(E_{eff}+V_{eff}[{\bf X}(i)]~)]^{-1/2}.
\label{time}
 \ee 
 
Hence, by adopting  the HJ formulation of Eq.~{\bf \ref{SHJ}}, it is possible to replace
 the  time discretization  with the discretization of the curvilinear abscissa $l$, which measures the
Euclidean distance covered in configuration space during the reaction~\cite{DRP0}.   This way,  the problem of the decoupling of time scales is bypassed.
 As a result, only about $10^2$ frames are usually sufficient to provide a convergent representation of a trajectory. 
On the other hand, the HJ formalism requires to perform an optimization in the space of reactive pathways of a functional, which can take complex values, which is in general
a complicated task. 

The DRP approach displays important similarities with the SDEL (Stochastic Difference Equation in Length)  method developed by Elber and co-workers~\cite{method4}. 
In particular, while the DRP is based on minimizing the effective HJ action $S_{DRP} = 1/\sqrt{D}\int dl \sqrt{V_{eff}[{\bf X}(l)]+E_{eff}}$, in   SDEL  the folding trajectories 
are obtained by {\it extremizing}   the {\it physical}  HJ action   $S_{SDEL} = \int dl \sqrt{U[{\bf X}(l)]-E}$, where $U(x)$ is the potential energy and $E$ is the conserved
 total mechanical energy.  
 
The SDEL algorithm is presently implemented in the MOIL software for molecular modeling~\cite{MOIL}. In this code, the protein folding trajectories are obtained by means of 
simulating annealing relaxation,  starting from an initial trial trajectory --see e.g. Ref. \cite{Elberproteins}---. 
The DRP algorithm is now implemented in DOLOMIT  (Dynamics Of muLti-scale mOdels of Molecular transITions), a program under development at the Interdisciplinary Laboratory for
 Computational Science (Trento). This code uses the MD-based algorithm described in the next section to identify the dominant path.
   
\subsection{Exploration of the Path Space}

The reliability of the DRP approach in investigating the protein folding transition crucially depends on the efficiency of the algorithm used to find optimum paths. 
In the analysis of conformational \cite{DRP2, QDRP2} or chemical~\cite{QDRP1} reactions of relatively small molecules, dominant paths can be found by  directly  
optimizing the HJ action in Eq.~{\bf \ref{SHJ}}, e.g.  using
simulated annealing or gradient-based methods. The DRP calculations for protein folding obtained this way 
have been extensively tested using reduced models in which the relevant degrees of freedom are
individual aminoacids and the energy landscape was relatively smooth~\cite{testDRP1, testDRP2}. Unfortunately, 
when moving from a coarse-grained to an atomistic description, a large amount of  frustration is introduced and the  energy surface of the model 
becomes much rougher. Consequently, the relaxation algorithms adopted in our previous calculations
 were found to provide a poor exploration of the space of folding paths in an all-atom calculation. 

In order to overcome this problem, we have used a  biased MD algorithm to  efficiently produce a large ensemble of paths, starting from a given 
denatured configuration and reaching the native state  \cite{ratchetMD2,ratchetMD3, method3}.
In particular, the  so-called \emph{ratchet-and-pawl MD} (rMD) algorithm \cite{method3} exploits the spontaneous fluctuations of the system along a specific  collective coordinate (CC),
 towards its native configuration.
This is done by introducing a time-dependent bias potential $V_R({\bf X},t)$, whose purpose is to make it very unlikely for the system to evolve back to previously 
visited values  of the CC. On the other hand, this bias exerts no work on the system when it spontaneously proceeds towards the native state. 
We emphasize that this approach is quite different from the one used in \emph{steered}-MD~\cite{steeredMD}, where an external force is continuously 
applied to the system, in order to drive it towards the desired state.

 Following the work of Ref. \cite{method3} we chose a CC $z(t)$,  which defines the distance between the contact map 
in the instantaneous configuration ${\bf X}(t)$ from the  contact
map in the native configuration ${\bf X}^{\text{native}}$. This way, the energy optimized native configuration has by definition $z=0$. On the other hand, a bias on $z$ does not force nor lock any 
specific contact, but only imposes 
a (quasi)~monotonic behavior of the \emph{total number} of native contacts.

In particular, the biasing potential introduced in Ref. \cite{method3} is defined as
\be
  V_R({\bf X}, t) = \left\{
    \begin{array}{lr r}
      \frac{k_R}{2} ( z[{\bf X} (t)] - z_m(t) )^2, &\text{for}&\quad z[{\bf X}(t)] > z_m(t)\\
      0, &\text{for}&\quad z[{\bf X}(t)] \leq z_m(t).\\
    \end{array}
  \right.\vspace{2mm}
  \label{VR}
\ee
In these equations, $z_m(t)$ is the minimum value assumed by the collective variable $z$ along the rMD trajectory, up to  time~$t$.
 
The value of the collective variable $z$ in the instantaneous configuration  $X(t)$ is defined as:
\begin{equation}
  z[X(t)] \equiv \sum_{i,j}^{N} [ C_{ij}[{\bf X}(t)] - C_{ij}({\bf X}^{\text{native}}) ]^2.
\end{equation}
The entries of the contact map C$_{ij}$ are chosen to  interpolate smoothly between 0 and 1, depending on the relative distance of the 
residues $i$ and $j$: 
\begin{equation}
  C_{ij}({\bf X}) = \frac { 1-(\frac{ r_{ij} }{ r_0 })^6 } { 1-(\frac{ r_{ij} }{ r_0 })^{10} },
\end{equation}
where r${_0}$=7.5~\AA~ is a fixed reference distance. 
The variable  $z_m$(t) is updated 
only when the system visits a configuration with a smaller value of the CC, i.e. any time $z[{\bf X}(t+\delta t)]< z_m(t)$.  

The value of the spring constant $k_R$ in the ratchet potential ---see Eq. {\bf \ref{VR}}--- controls the amount of bias introduced by the ratchet algorithm. 
At very low values of $k_R$, the rMD trajectories are minimally biased. However, in this limit the system performs the 
first folding transition on typical time intervals  comparable with the folding time $t\sim 1/k_f$ ---where
$k_f$ is the folding rate---. 
On the other hand, the DRP method is based on the transition probability given Eq. {\bf \ref{Pf}}, where the time interval $t$ is of 
the order of the transition path time $t_{TPT}\ll 1/k_f$. 
Hence, DRP simulations based on ratchet simulations in which $k_R$ is chosen too small are computationally very expensive, since most paths do not reach the 
native state within the time interval $t$ entering   Eq. {\bf \ref{Pf}}, hence do not contribute to the transition probability. 

In the opposite high $k_R$ limit,  the bias force becomes comparable with the physical internal forces acting on the atoms and the dynamics is affected by a significant bias. In this regime, if the ratcheting coordinate is not optimal, the system is driven into relatively large free energy regions. The unbiased statistical weight given by Eq. {\bf \ref{prob}} 
penalizes these trajectories. 
In the extremely high $k_R$ limit,  the bias towards forming native contacts is very large, and breaking a native contact which is present in a (partially) misfolded configuration becomes extremely unlikely. As a result, also  rMD simulations in which $k_R$ is chosen too large have a low folding yield, since the trajectories  have a large probability to be trapped in misfolded states.
  
This algorithm allows to efficiently generate a large number of trajectories starting from the same configuration and reaching the native state, hence it can be used to explore the 
folding path space. Eq. {\bf \ref{prob}} provides a rigorous way to score such trial trajectories, i.e. 
to evaluate the probability for each of them to be realized in an \emph{unbiased} overdamped Langevin dynamics simulation.
In particular, the best estimate for the dominant folding pathway is the one with the smallest Onsager-Machlup action. 
The path may then be used as a starting point for a further refinement based on a local relaxation of the HJ action given by 
Eq.~{\bf \ref{SHJ}}, performed by means of the optimization algorithms described in our previous work (see e.g. Ref. \cite{QDRP2, QDRP3}).

The second refinement step is computationally very  expensive, requiring several thousands of CPU hours for each dominant trajectory. However,
 by performing a number of test simulations, we have found that it produces only very small  rearrangements of the chain, mostly  filtering out small thermal fluctuations
(see e.g. Fig.\ref{Fig4}).

Some comments on the minimization procedure described above are in order:  first, we emphasize that since the dynamics generated by the rMD algorithm has infinite memory, it
violates the detailed-balance condition and  the time intervals computed from the biased trajectories have no direct physical meaning. 
On the other hand, once a minimum of the HJ action has been identified,  it is in principle possible to reconstruct the physical time at which each frame is visited, by means of
 Eq.~{\bf \ref{time}}.

As long as one is concerned mostly with the global qualitative aspects of the folding mechanism, the expensive refinement session of the DRP calculation may be dropped. 
This  allows  us to reduce  the total computational time required to perform the  analysis by several orders of magnitude. 
On the other hand, we are not in a condition to provide a reliable estimate of the 
time interval that the protein takes to complete a folding reaction. 

We stress that the present DRP approach is expected to give unreliable results if the sampling of the trial trajectories is too limited or if the CC 
 adopted in the bias potential of the rMD algorithm was a bad reaction coordinate. 
A number of recent studies have shown that the total number of native contacts represents reasonable reaction coordinate for the folding of small globular proteins
~\cite{ reactioncoord2, reactioncoord1,reactioncoord3}.

\begin{table*}[th]
\caption{Summary of the simulation details}
\begin{tabular*}{\hsize}
{@{\extracolsep{\fill}}rrrrr}
\hline
\multicolumn1c{Unfold. MD steps} &
\multicolumn1c{Therm. MD steps } &  
\multicolumn1c{Fold. rMD steps} &  
\multicolumn1c{Denatured} &
\multicolumn1c{Rejected} 
 \cr 
\multicolumn1c{(T= 1600 K)} &
\multicolumn1c{(T= 300 K)} &
\multicolumn1c{ (T= 300 K)} &
\multicolumn1c{initial conditions} &
 \cr
\multicolumn1c{$50 \times 10^3$\tablenote{In all MD and rMD simulations the integration time step has been chosen to be 1 fs.}} & 
\multicolumn1c{$100 \times 10^3$} & 
\multicolumn1c{$50 \times 10^3$} & 
\multicolumn1c{44} & \multicolumn1c{18}
\cr
\hline
\end{tabular*}
\label{table}
\end{table*}

\subsection{Computational Procedures and Simulation Details}

The atomistic DRP calculations were performed using DOLOMIT. The code calls a 
librarized version of GROMACS 4.5.2~\cite{GRO4} to calculate the molecular potential energy and its gradient. 

We defined the native state as the set of conformations with a RMSD to the crystal structure of the $C_\alpha$  in the hairpins smaller than 3.5 \AA. 
A configuration was considered
denatured if the RMSD to native of both hairpins was larger than 6\AA.  
The stability of the native state within the present force field was checked by running 12 unbiased 2 ns-long MD simulations at the room temperature (300 K). In all such trajectories
the protein remained in the native state. We then generated 44 independent initial conditions, by running a 50~ps MD at 1600~K, starting from the 
energy minimized native state, followed by a 100 ps relaxation at 300~K. The  time step employed in all the simulations was 1 fs.
From each of the 24  starting configurations we performed 96 independent runs, each consisting of 50000 rMD steps. For each of the remaining 20 initial conditions, the number of trial paths was limited to 48.  

In our simulations, we have chosen the ratchet spring constant
  $k_R=0.02$~kcal/mol, which represents a reasonable compromise between keeping a high computational efficiency for the path exploration
and introducing a small bias. Indeed, using this value, the modulus of the biasing force is always at least one order of magnitude smaller
than the norm of internal forces.

We observed that  5 of  the 44 initial conditions did not correspond to denatured states,  hence they were rejected. 
In addition, in 13 of the remaining 39 sets, more than $80\%$  of the trial trajectories did not reach the native state within the simulation time. In  these cases 
the exploration of the  path space was limited to very few trial trajectories, so the corresponding dominant paths were discarded. 
For the remaining 26 sets of paths, we identified the most probable by computing the OM action of Eq. {\bf \ref{prob}}. 
The complete set of atomistic simulations required less than two days of  calculation on 48 CPU's.

It is important to study to what extent DRP results obtained this way depend on the value of the bias constant $k_R$ adopted in the rMD 
  simulations. 
  In Fig. \ref{Fig5} we plot the dominant reaction pathway obtained starting from the same initial condition, using different values of $k_R$ which span over almost two orders of 
  magnitude.    We see that in most  simulations the folding occurs through the same qualitative pathway, in which the first hairpin forms before the second begins to fold. Only
 in one case ---for a low value of the coupling constant $k_R$--- we find that the protein travels across the denatured state before taking a different pathway to the native state,
   in which the order of formation of the hairpins is reversed.  
 Such a trajectory spends a much longer fraction of rMD steps in exploring the denatured state and initiates the transition from a very different configuration. 
  
  It is interesting to note that the most probable path turns out not to be the one with the  lowest ratchet constant.
  In particular, the trajectory taking the second pathway (labelled with $K_1$ in Fig. \ref{Fig5}) is among those with the lowest statistical weight. This is because the OM action
   tends to penalize paths which travel long Euclidean distances in configuration space. This result
  illustrates that choosing a low ratchet constant $k_R$ produces an enhanced exploration of the denatured state, but does not necessarily lead to a more
   efficient identification of the most probable trajectories connecting given boundary condition at fixed times. 
  
Once a dominant path has been found, it is relatively straightforward to identify
the configuration which belongs to the transition state ensemble. This can be done by finding the frame ${\bf X}_{TS}$ in the trajectory such that 
 the probability to reach the native state is equal to that of going  back to the denatured configuration~\cite{DRP2}:
 \be
 \label{TS}
\frac{\text{Prob}[{\bf X}_{TS}\to \text{Unfolded}]}{\text{Prob}[{\bf X}_{TS}\to \text{Native}]} \simeq \frac{e^{-S_{OM}({\bf X}_{TS}\to {\bf X}_D)}}{e^{-S_{OM}({\bf X}_{TS}\to {\bf X}_N)}}=1
\ee
In this equation ${\bf X}_N$ and ${\bf X}_D$ are the first native and denatured configurations visited along the dominant path, starting from ${\bf X}_{TS}$.
In order to locate them, we need to only take into account the ``reactive'' part of the path, that is the one which leaves the denatured state and, without recrossing, 
goes straight to the native. To satisfy this requirement, we considered the total RMSD versus frame index curve. The typical trend of this curve for most of the dominant trajectories  is 
shown in the lower panel of Fig.~\ref{Fig6}: it consists
in an initial plateau, followed by a rather steep fall, and then by another flat region, where
the system oscillates in the native state. The reactive part of the path was identified with the region of steep fall in this curve. In particular,
the beginning of the transition was set to the frame at which the derivative of the total RMSD curve changes sign, from positive to negative.

The simulations of the equilibrium properties of the WW-domain in the coarse-grained models were
performed using a Monte Carlo algorithm based on a combination of Cartesian, crankshaft~\cite{crankshaftMC} and pivot~\cite{pivot} moves. 
The free energy as a function of an arbitrary set of reaction coordinates (potential of mean force) was obtained from the frequency histogram 
calculated from long MC trajectories. 

\section{Results and Discussion}

In the upper panel of Fig.\ref{Fig6} we show our set of atomistic  dominant folding trajectories, projected onto the plane defined by the
 RMSD to the native structure 
of the C$_{\alpha}$ atoms in residues $8-23$ (hairpin~1) and $17-30$ (hairpin~2).

Two distinct folding pathways which differ by the order of formation of the hairpins can be clearly identified: in about half of the computed dominant
 folding pathways hairpin~1 consistently folds before hairpin 2. 
 In this  channel, we find the transition state is located at the ``turn" of the paths, i.e. is formed by  configurations in which the hairpin 1 is folded 
while hairpin 2 is largely  unstructured (see pathway 1 in  Fig. \ref{Fig2}). This is the mechanism predominantly  found in the simulation of Ref.~\cite{theWW5}, performed using the same force field, albeit in explicit solvent.  

In about half of the computed dominant paths, we observe that the two hairpins form in the reversed order. In this channel, the transition state is formed by the 
configurations in which hairpin 2 is 
folded, while hairpin 1 is unstructured (see pathway 2 in Fig.~\ref{Fig2}).

An important point to emphasize is that this clean two-channel folding mechanism is a prediction of the DRP approach which could not be obtained if we analyzed directly 
the full set of (biased)  trajectories obtained from the rMD algorithm, without scoring their probability. 
For example, Fig.~\ref{Fig7} shows that not all the rMD trajectories computed starting from a given initial condition follow one of the two folding pathways discussed above.  
Indeed,   many of them
 involve a simultaneous formation of native contacts in both hairpins. A clear prediction of the DRP formalism is that folding events in which the hairpins form simultaneously are much less
frequent than those in which the two secondary structures forms in sequence. 

Another result emerging from our DRP calculation is the existence of a correlation between the structure of the initial conditions from which the transition is initiated 
and the pathway taken to fold: if at the beginning of the transition 
the first hairpin has a RMSD smaller than the second hairpin, then the first pathway is most likely chosen. 
In the opposite case, i.e. when the second hairpin has a smaller RMSD to native than the first, then the second pathway is generally preferred. 

In order to further support  these results and gain insight into the folding mechanism, we have performed simulations in an entirely different approach, i.e. by computing equilibrium properties
using the coarse-grained models described in the Methods section.  In Fig. \ref{Fig3} we show the free energy landscape at the 300~K, as a function of the RMSD 
to native of the two hairpins  for the two models,  which differ by the presence of non-native interactions. 
In both cases, we observe the existence of two valleys in the free energy landscape, which  correspond to the two folding pathways discussed above.  Remarkably, the same structure for this
free energy map was obtained by Ferrara and co-workers for the 20-residue peptide  beta3s ---which shares the same native topology of WW domains \cite{ferrara}---
  by means of equilibrium atomistic simulations based on the CHARMM force field, in implicit solvent. 
 
We emphasize that this set of  qualitatively consistent results has been obtained using different algorithms (in equilibrium and nonequilibrium conditions) and different theoretical models 
(atomistic and coarse-grained). The fact that models with and without non-native interactions give very similar  free energy landscapes suggests 
 that the structure of the two folding pathways of WW domains is mostly shaped by native interactions.  

\subsection{Comparison with experimental data on folding kinetics}

In general, all experimental data on folding kinetics  of WW domains indicate that the formation of the first hairpin is the main rate limiting step~\cite{expWW1, expWW2, expWW3}. 
In particular, the $\phi$-values measured by J\"ager and co-workers display a clear peak in the region associated with hairpin 1, but  significant
 $\phi$--values were reported also for residues in the sequence region relative to hairpin 2~\cite{expWW3}. This fact indicates that the folding of the latter structure has some rate limiting effect.   
In addition, it was found that the $\phi$--values in the region of the second hairpin grow with temperature, while  those  in the region of the first hairpin  decrease. 
This implies that, at higher and higher temperatures, the second hairpin plays an increasing role in the folding mechanism.

An analysis based on $\phi-$values alone does not permit to fully characterize the folding mechanism. In particular, it cannot distinguish between a single-channel folding 
mechanism in which native contacts in the 
two hairpins form simultaneously and a multiple-channel folding mechanism  in which the reaction rate in each channel is limited by the folding of one of the hairpins.
 In Ref. \cite{weikl} Weikl has shown that the full body of existing $\phi$--value data (taken from Ref.s \cite{expWW2, expWW3})  can be consistently and quantitatively explained by a simple kinetic model in which the folding of WW domains occurs through alternative  channels,
 which correspond  to the two pathways found in our DRP simulations.  From a global fit of the experimental data, the author concluded that the relative probability of the
 first folding pathway for FBP and Pin 1 WW domain  are $77\% \pm5\%$ and $67 \pm 5\%$, respectively. 
 
\begin{figure}
\includegraphics[width=8 cm]{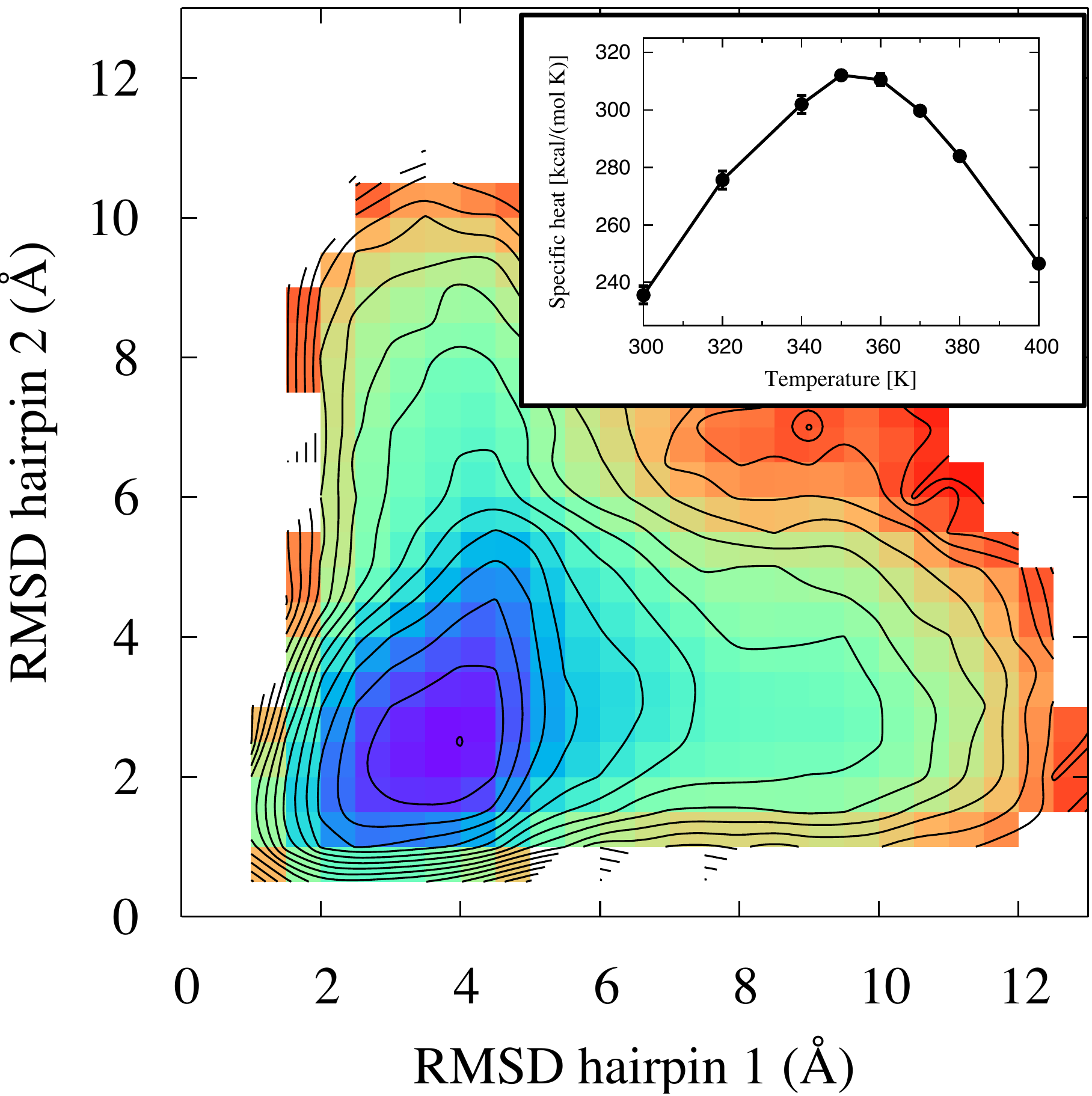}\\
\includegraphics[width=8 cm]{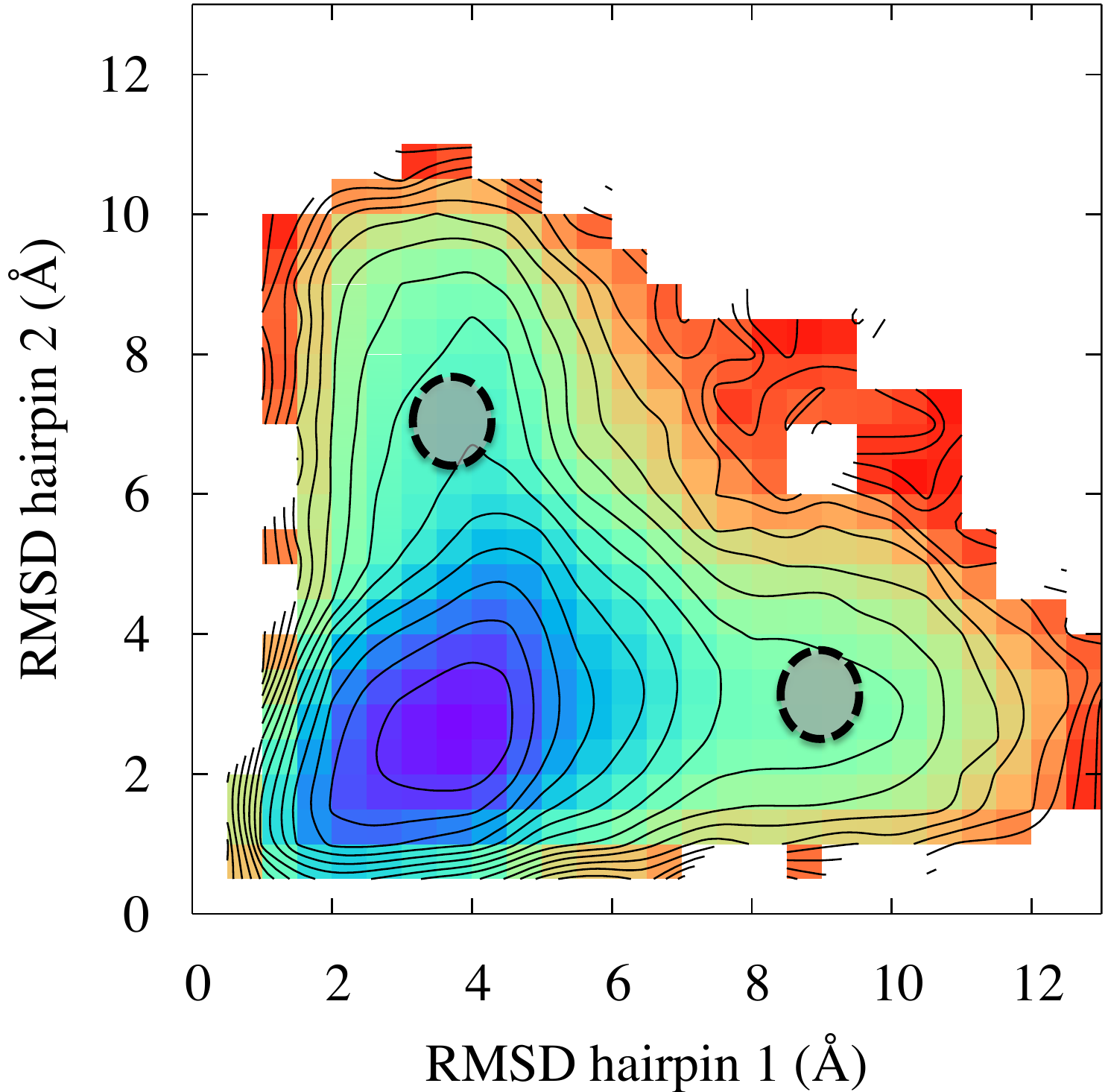}
\caption{\bf 
The free energy at $T=300 K$ as a function of the RMSD to native of the two hairpins, obtained from the Monte Carlo simulations in two coarse-grained models, 
described in the Method section. In the upper panel the model accounts for both native and non-native interaction, in the lower panel the model contains only 
native interactions. In the insert of the upper panel, we show the corresponding plot of the specific heat. Both models predict the existence of two distinct folding paths. The two shaded regions in the lower panel,  with coordinates in the range  (3.5 $\pm$ 0.5 \AA, 7 $\pm$ 0.5 \AA) and 
(9$\pm$ 0.5 \AA, = 3.0 $\pm$ 0.5 \AA) are identified with the average location of the transition states obtained with DRP simulations. 
The average of the free energy in these region are used to estimate the relative rate of the two reaction channels.}
\label{Fig3}
\end{figure}

\begin{figure}
\includegraphics[width=10 cm]{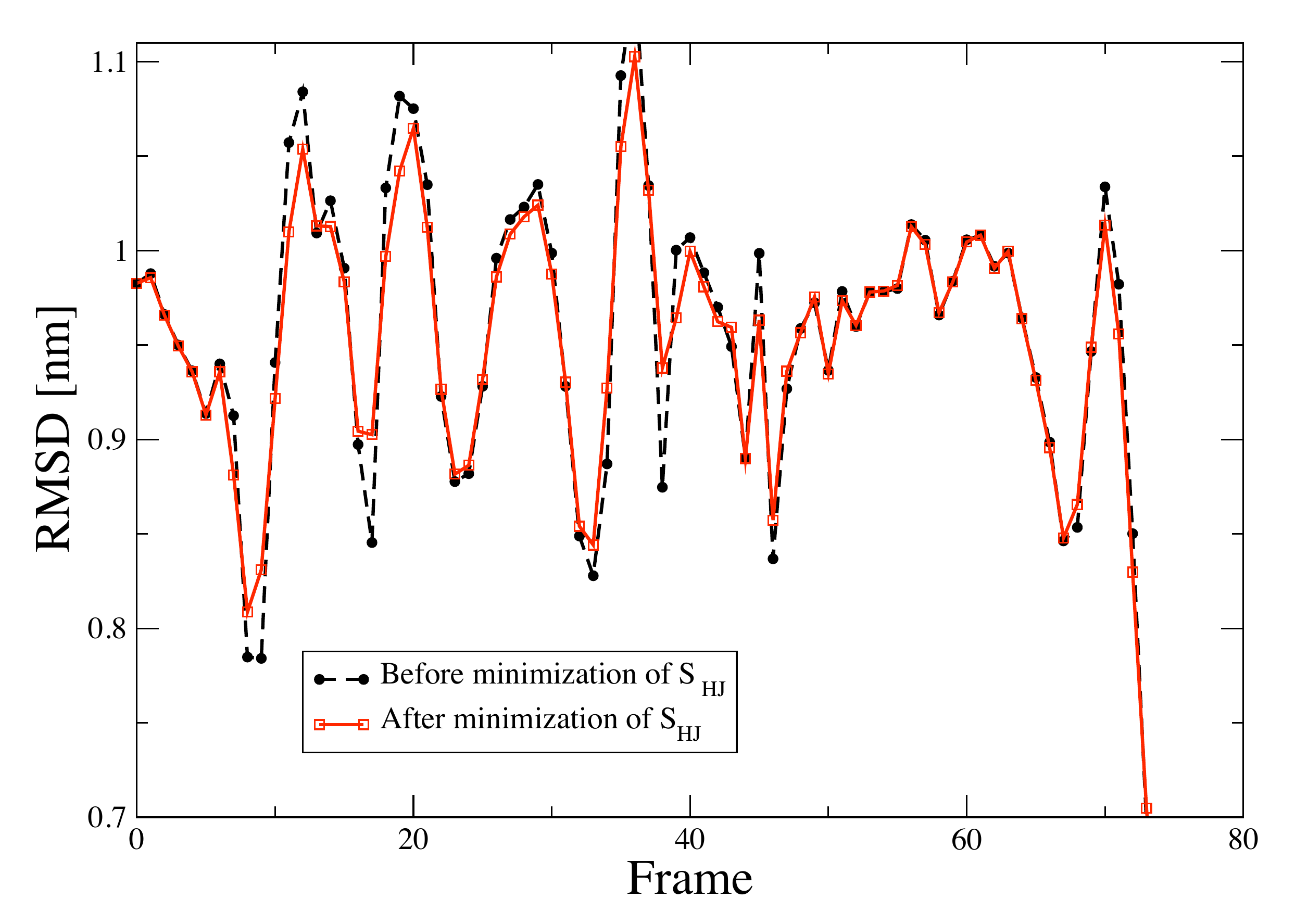}
\caption{\bf The effect of the minimization of the HJ functional {\bf \ref{SHJ}} on a trial path obtained from the rMD algorithm on the RMSD to native
of the $C_\alpha$ atoms in the backbone. This second refinement step does not change the qualitative structure of the dynamics. }
\label{Fig4}
\end{figure}

\begin{figure}
\includegraphics[ width=12 cm]{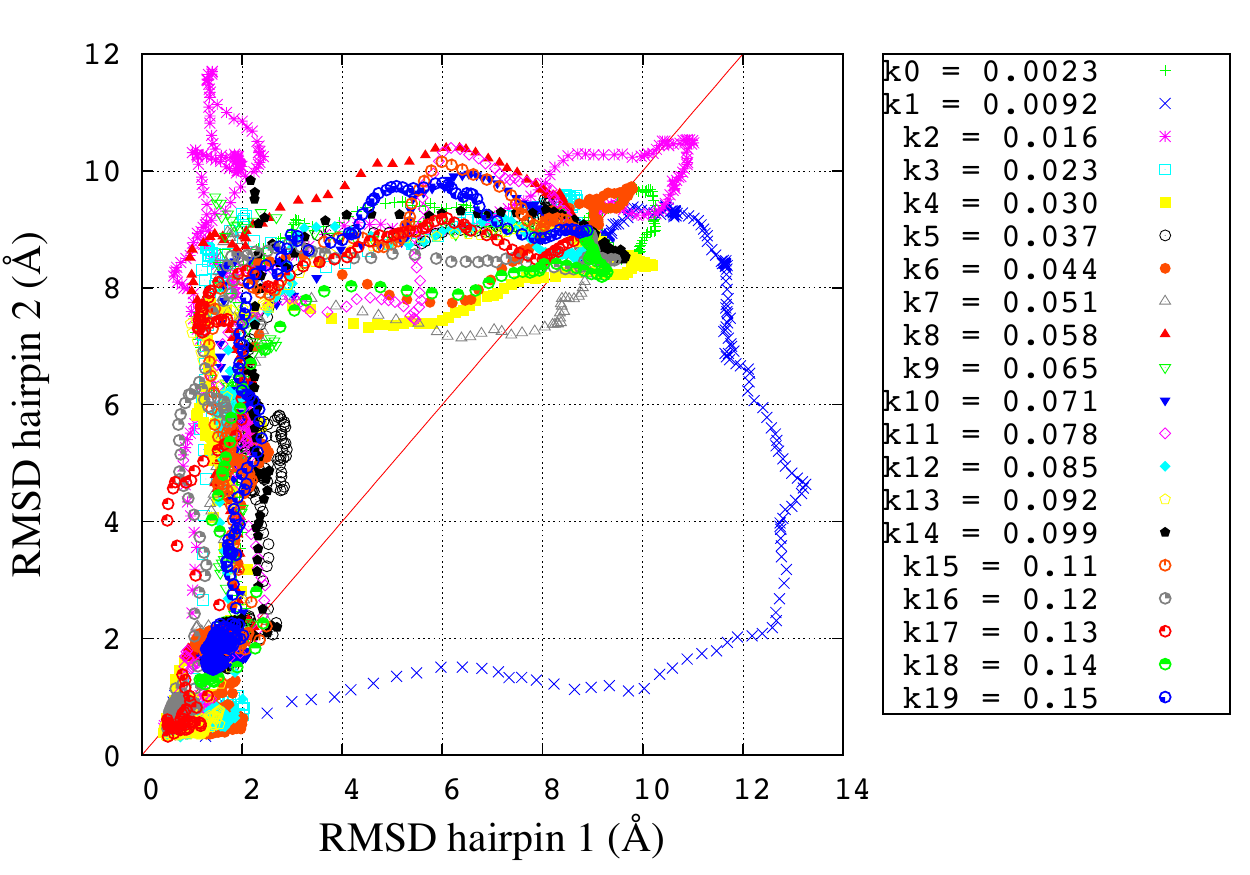}
\caption{\bf Dominant pathways obtained with different values of the ratchet force, starting from the same initial condition.
The values of $k_R$ are given in units of kcal/mol.   }
\label{Fig5}
\end{figure}

\begin{figure}
\includegraphics[width=13 cm]{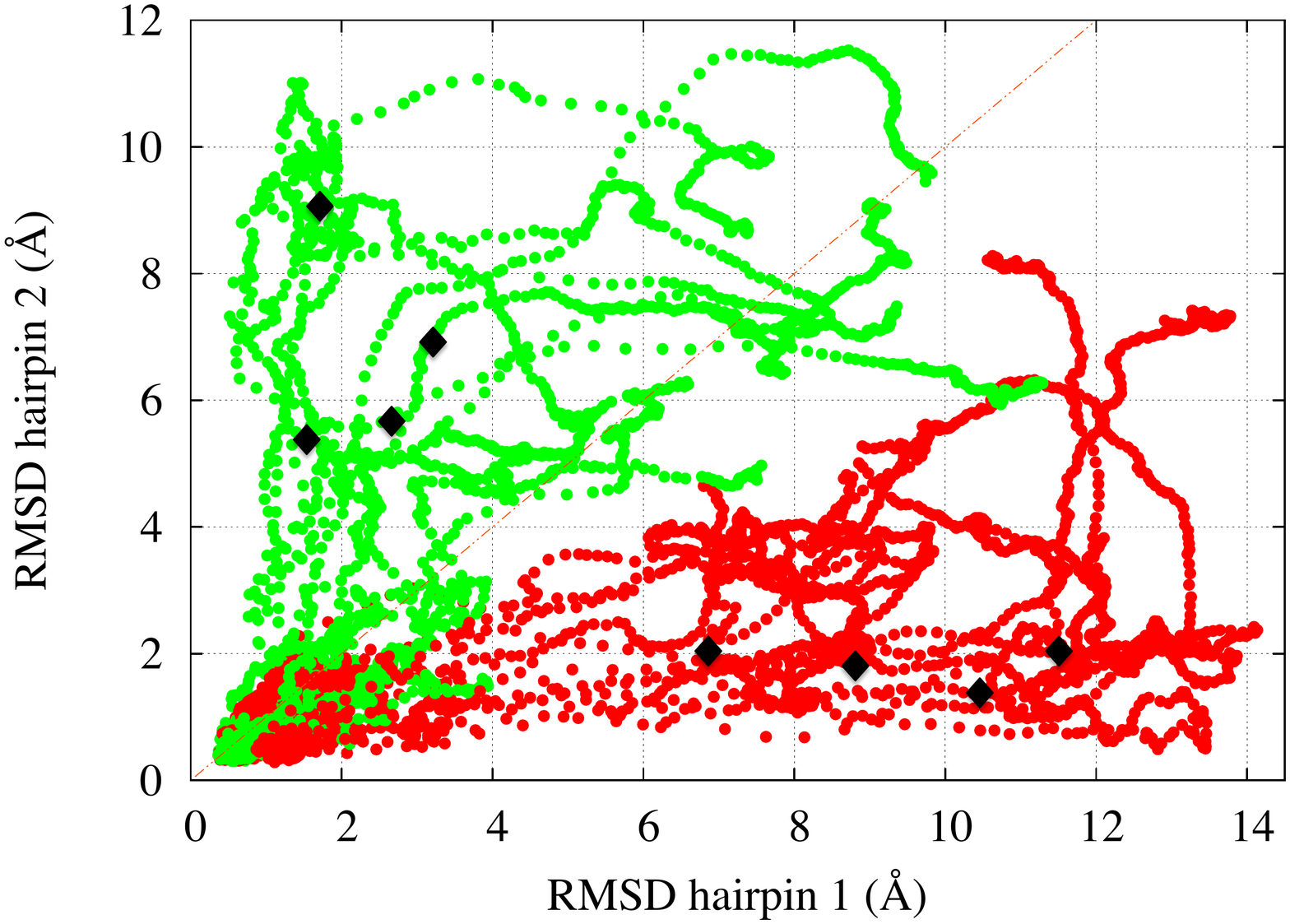}\\
\includegraphics[width=13 cm]{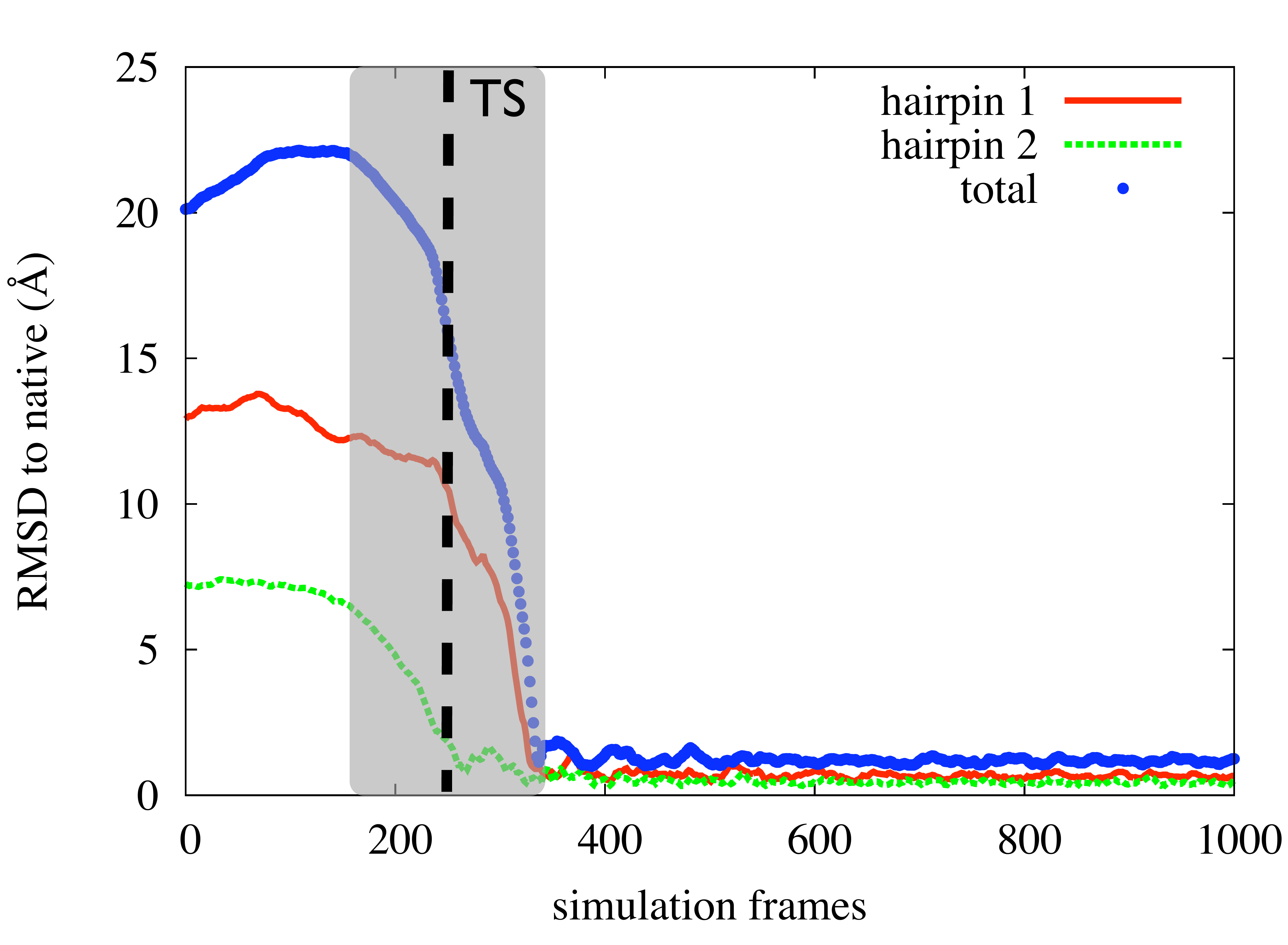}\\
\caption{\bf 
Upper panel: the set of dominant folding paths for Fip35, obtained from atomistic DRP simulations, projected on the plane defined by the RMSD of the two hairpins to the corresponding native structures. 
 The dark spots represent a few typical configurations in the two transition states, evaluated by the requesting a probability 1/2 to reach
 the native state. Lower panel: evolution of the RMSD to native of the full protein and of the hairpins along a dominant trajectory. The shaded area is the reactive region and the 
 dashed line identifies the transition state configuration obtained according to Eq. {\bf \ref{TS}}. 
 }
\label{Fig6}
\end{figure}

\begin{figure}
\begin{center}
\includegraphics[width=10 cm]{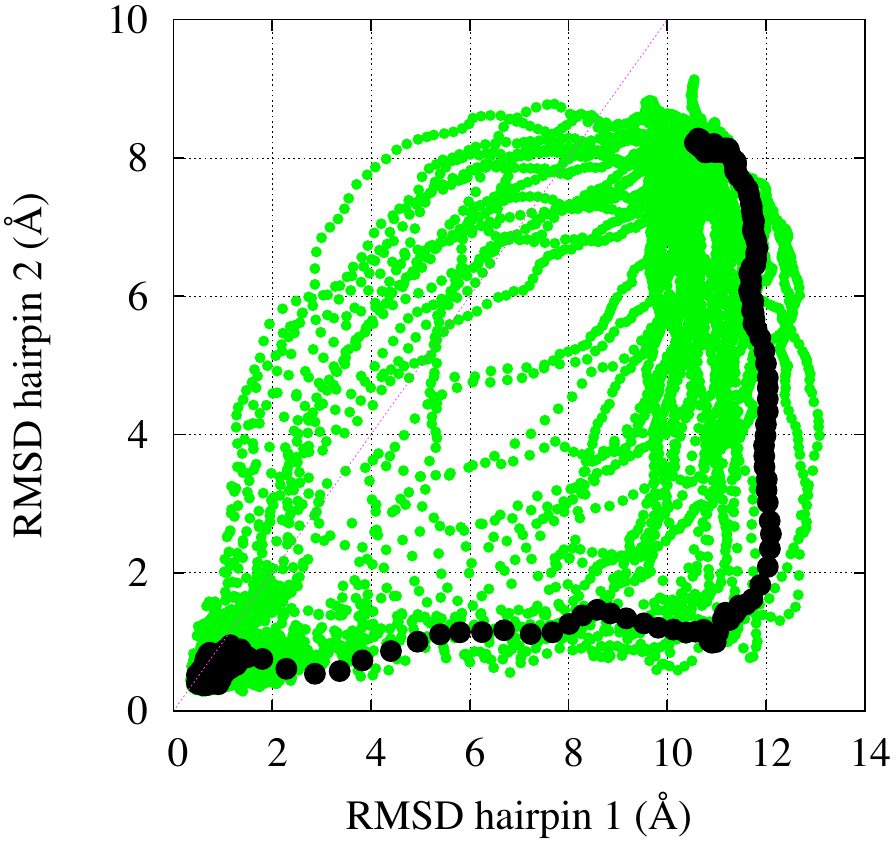}\\
\includegraphics[width=10 cm]{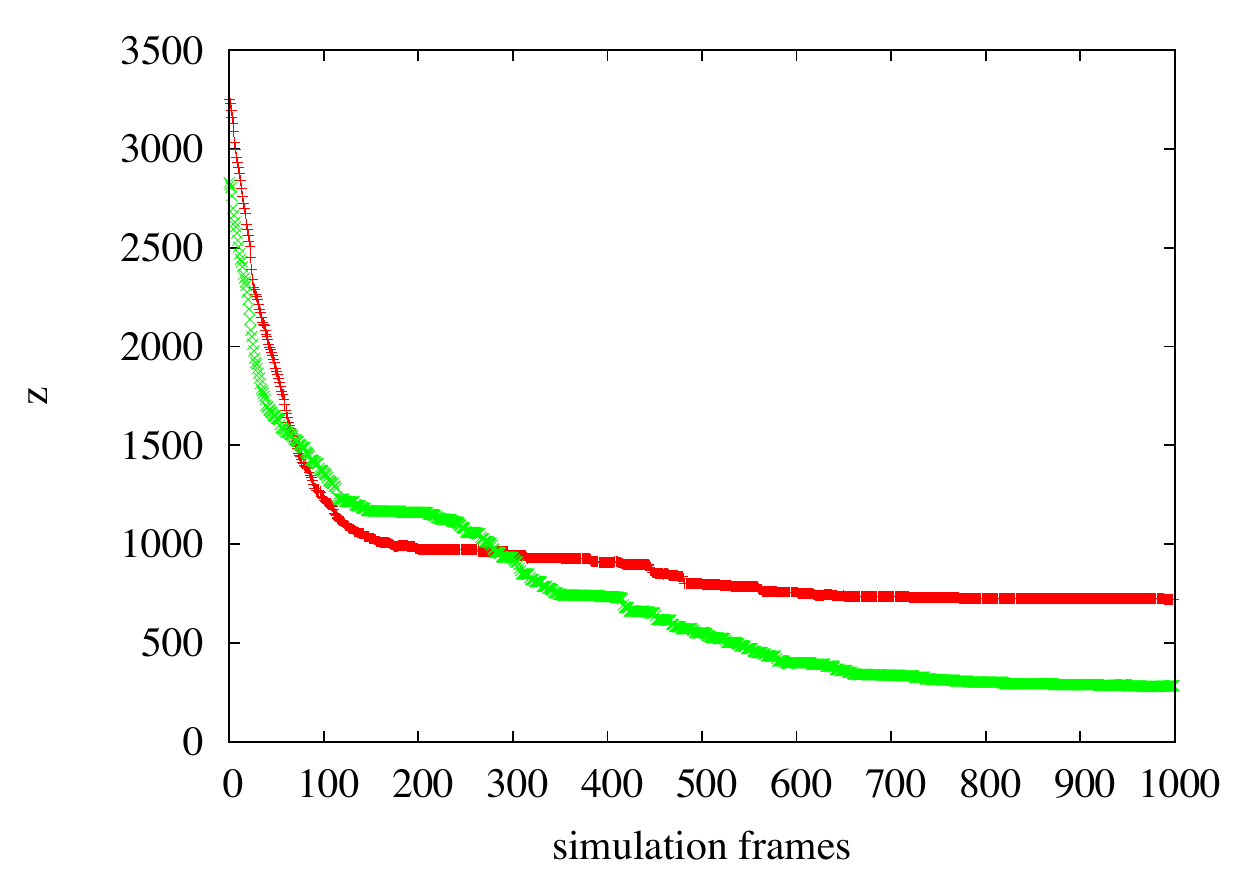}
\caption{\bf 
Upper panel: The set of trial paths  connecting a given denatured configuration to the native state, used in the search for  the dominant paths, projected on the plane defined by the RMSD to native of the two hairpins. The darker path is the selected
dominant reaction pathways. Lower panel: the  evolution of the biasing coordinate, in two typical folding trajectories of this set.}
\label{Fig7}
\end{center}
\end{figure}

Let us now discuss the relative statistical weight of the two folding pathways. To this goal we need to estimate and compare
 the reaction rates in the two channels.   The formalism  for  evaluating reaction rates in the  DRP approach was developed in detail in Ref. {\bf \cite{DRP4}}, 
 where it was shown that  this method reproduces Kramers theory in the low-temperature regime.  Applying that formalism, the ratio of the folding rates in the two channels reads
\be
\label{rate}
\frac{k_1}{k_2} \simeq \frac{\kappa^1_0}{\kappa^2_0}~e^{-\beta~(G_{TS_1}-G_{TS_2})},
\ee  
where the label 1 (2) identifies the channel in which hairpin  1 (2) folds first. 
In  Eq.~{\bf \ref{rate}}, the exponent contains the difference of the free-energies of the two transition states, defined from the dominant trajectories according to the commitment analysis
described in the Methods section and in Ref.~ \cite{DRP2}. 
In particular, one has 
\be
\label{Gi}
e^{-\beta G(TS_i)} \equiv \int d{\bf X}~ e^{-\beta U({\bf X})} \delta\left[({\bf X} -{\bf X}_{TS}^i) \cdot \hat n_{TS^i}\right]\qquad (i=1,2),
\ee
where ${\bf X}^i_{TS}$ is a point of the  transition state which is visited by a typical dominant path in  the $i-$th reaction channel and $\hat n_{TS^i}$ is a
 versor tangent to the dominant path  at ${\bf X}_{TS}^i$. Using Eq.~{\bf \ref{TS}} to identify the transition states, we have found that the two partition functions defined in Eq.~{\bf\ref{Gi}} are dominated  by configurations in which one of the hairpin is 
 fully formed while the other is still completely unstructured (see Fig. \ref{Fig6}). The average location of the computed transition states in the plane defined by the
 RMSD to native of the two hairpins is highlighted in the lower panel of  Fig. \ref{Fig3} and can be identified with the endpoints of the two low free energy valleys.

 The coefficients $\kappa_0^1$ and $\kappa_0^2$ in the prefactor of Eq.~{\bf \ref{rate}} are defined in terms of quantities which can be calculated from the dominant  paths --- see Ref.\cite{DRP4} for details---. These terms
   estimate the average flux of reactive trajectories 
across the iso-committor dividing surface, including the contributions from  small thermal fluctuations around the dominant paths. Unfortunately, evaluating $\kappa_0^1$ and $\kappa_0^2$
necessarily requires to perform the computationally expensive local optimization of the HJ action. 
However, if the reaction is thermally activated, the ratio of rates $k_1/k_2$ is mostly controlled by the exponential contribution. Hence, we can consider the Arrhenius 
approximation
\be\label{ratearrhenius}
\frac{k_1}{k_2} \simeq e^{-\beta~(G_{TS_1}-G_{TS_2})}.
\ee
We emphasize that in this approach the DRP information about the nonequilibrium reactive dynamics is used to define the two transition states. 
On the other hand, the numerical value of the free energy difference may be obtained from equilibrium techniques, e.g.  by sampling the integrals in Eq.~{\bf \ref{Gi}} by means of  computationally very expensive umbrella sampling  or
  meta-dynamics~{\bf \cite{Metadinamica}} atomistic calculations. 
  
In this first exploratory application of the DRP formalism to a realistic protein folding reaction  we choose to perform a much rougher estimate which relies on two main approximations. First, we identify the difference $(G_{TS_1}-G_{TS_2})$ in Eq.~{\bf \ref{ratearrhenius}} with the difference of the free energy in the two shaded regions
 of the energy landscape shown in Fig. \ref{Fig3}. The centers of these regions represent the average location of the configurations in the two transition states TS1 and TS2 obtained from DRP simulations, projected onto the plane selected by the RMSD to native of the two hairpins. The sizes of the shaded area represents the errors on the average location of the transition states on this plane, estimated from the standard deviation.
The second assumption of our model is 
that such a  free energy difference is driven by the balance between energy gain and entropy loss associated to the formation of  \emph{native} contacts in the two hairpins.  This native-centric  standpoint is supported in part by the fact that
free energy landscapes computed in different models with and without non-native interactions are found to be very similar, as it is clear from comparing the panels in Fig.~\ref{Fig3}. 
Hence, to estimate 
 $G_{TS_1}-G_{TS_2}$ we used  the G$\overline{o}$-type model described in the Methods section. 
 It is important to emphasize that we are not computing the rate directly from a transition state theory formulated in the coarse-grained model,
 but we are using it only to estimate a free energy difference.
  
This way, we obtained an estimate $k_1/k_2 \simeq 2.3$ which corresponds to a relative weight of the first folding channel of  $70\%$ and $30\%$. 
We stress that,  such a simple calculation should be considered only a rough estimate. It indicates that the two channels have more or less comparable weight and that the first channel is the most
probable, in qualitative agreement with experimental results and with the simulations of Shaw and co-workers.

This simple scheme enables us to address the question of the dependence of the relative weights of the two channels on the temperature. 
Repeating the calculation at a higher temperature of  $380$~K --- assuming that the structure of the transition states is not significantly modified--- we find $k_1/k_2 \simeq 1.6$, which corresponds to a branching ratio
 of channel 1 of about $60\%$.  Hence,  the rate limiting role of the second channel grows with temperature, in qualitative agreement with experimental kinetic data. 

This fact can be understood as follows. The folding of one of the hairpins generates an entropy loss  proportional to the number $n$  of native contacts formed.
More precisely, assuming that the protein in the unfolded state is completely denatured then the entropy loss in the TS of the first (second) channel is $\Delta S_{TS1 (TS2)}= n_1 (n_2) s$, where $s$ is the entropy loss for locking  one residue in its native conformation, while $n_1=15$ and $n_2=13$ are the numbers of residues in the two hairpins of Fip35.
The transition state in the first channel involves forming a longer hairpin, hence reaching it involves a larger entropy loss (but also larger gain of native energy). 
The role of the entropy loss relative to the energy gain in forming the hairpins grows with temperature, hence
 disfavoring the first folding channel relative to the second.

 \section{Conclusions}

In this work we have studied the folding mechanism of the WW domain Fip35
 in equilibrium  and nonequilibrium conditions, using an atomistically detailed force field and coarse-grained models based on knowledge-based potentials.
   
The obtained folding trajectories are not heterogeneous, but rather suggest that the folding proceeds through two dominant channels, defined by a hierarchical order of hairpin formation.
In the most  probable channel at room temperature, the first hairpin is almost fully formed before the second hairpin starts to fold.
In the alternative route, the order of hairpin formation is reversed.  
Our simulations also suggest that the choice of the folding pathway is correlated with the  structure of the  denatured configuration from which the peptide initiates the  folding reaction. 
Our results are compatible with both Weikl's analysis  of kinetic data and Krivov's analysis of equilibrium MD simulations.   

The most important result of this work is to show that, using the DRP approach, it is possible to characterize at least the main qualitative aspects of the folding mechanism 
at an extremely modest computational cost, in the range of few hundreds of CPU hours.  Such a level of computational efficiency opens the door to the investigation of the folding pathways of a large number of single-domain proteins, with sizes significantly larger than that of the small domain studied in the present work.  

\begin{acknowledgments}
The DRP approach was developed in collaboration  with H. Orland, M. Sega and F. Pederiva. 
We thank G. Tiana and C. Camilloni for sharing important details of their implementation of the ratched-MD algorithm and S. Piana-Agostinetti for providing us with details on the results
of the  MD simulations of Ref. \cite{theWW5}.  

All the authors are members of the Interdisciplinary Laboratory for Computational Science (LISC), a joint venture of Trento University and the Bruno Kessler Foundation. 

SaB  and T\v{S} are supported by the Provincia Autonoma di Trento, through the AuroraScience project. 
Simulations were performed on the AURORA supercomputer located at the LISC (Trento) and partially on the TITANE cluster, which was kindly made available by the IPhT of  CEA-Saclay. 

\end{acknowledgments}


%
%

\end{document}